# A Rational Model of Large-Scale Motion in Turbulence

*Wennan ZOU*

*Institute for Advanced Study/Institute of Engineering Mechanics,*
*Nanchang University, Nanchang, Jiangxi, China 330031*
E-mail address: zouwn@ncu.edu.cn

***Abstract*—A rational theory is proposed to describe the large-scale motion in turbulence. The fluid element with inner orientational structures is proposed to be the building block of fluid dynamics. The variance of the orientational structures then constitutes new fields suitable to describe the vortex state in turbulence. When the fluid element is assumed to be an open subsystem, the differentiable manifold description of turbulence ought to be set up, and the complete fluid dynamics can be deduced from a variational calculus on the constructed Lagrangian dissipation energy density. The derived dynamical equations indicate that the vortex evolution is naturally related with the angular momentum balance.**

***Keywords*-Turbulence; Fluid element; Vortex field; Lagrangian dissipation energy density.**

## I. INTRODUCTION

Most fluid flows in engineering applications and nature are turbulent, that means the transition to turbulence in fluid flow is an everyday experience of ours. But turbulence is still a great challenge to human intelligence [1]. The most accepted theory of fluid flow in hydrodynamics community is the Navier-Stokes (N-S) equations. Many people believe that the real turbulence as a macroscopic phenomenon has to be the general solution of the N-S equations [2], and easy to arrive at a conclusion that turbulence is hopelessly difficult to predict — turbulence is beyond analytics [3], therein a basic fact is that nowadays a thorough mathematical understanding of the N-S equations is still lack. But some people insist on the understanding of the basic physics in turbulent flows, and recognize that there is no satisfactory theoretical framework to handle turbulence, say, the formation and evolution of structures in turbulence have not been described effectively [2]. In front of the intricacy and complexity of turbulence, we are inapprehensive and diffident to the call for new ideas and means. This paper is a decisive attempt at building a new theory of turbulence.

Conventionally, when turbulence appears in flow at high Reynolds (Re) numbers, the sensitive volume used to measure the fluid velocity seems to be smaller and smaller because of the turbulent fluctuation. Theoretically there is no limit to the smallest volume if the Re number has no limit, though it should be large comparing with the microscopic scale [4]. A terrible mess in turbulence research is said to be the number of sensitive volume calculated from the Re number is by a long way larger than the present computational ability. On the other hand, the ensemble average equations derived from the idea as proposed by Osbourne Reynolds more than a century ago, have to encounter the problem of closure. In this paper, we will adopt an absolute different way to take account of this process.

If we have to recognize both the universality of small-scale motion and the significance of large-scale coherent motion, there should exist a middle scale range of the sensitive volume in fluid flow, larger than the scale accounting for the small-scale motion and smaller than the scale accounting for the coherent motion. Then the kinematic quantities measured within this volume would cover the contributions of two kinds of fluctuant motion besides the mean velocity: the small-scale motion results into the so-called micro-rotation, and the topological configuration of fluid, which makes the fluid particles naturally move in a curved way, represents the structures embeding in the coherent motion. Basing on this point of view, a new topological framework will be proposed to describe the intricate structures in turbulence, in which some new field variables are introduced to indicate the vortex evolution.

The paper is constructed as follows. We first introduce the fundamental features of our model and the new variables used to embody them, and then clarify the kinematical features of motion fields and propose the coupling mechanisms between different field variables. Basing upon these ideas, we construct the dissipation energy density of fluid system in order to derive the dynamical equations from the variational calculus. Finally, we discuss the physical meaning of the derived governing equations.

## II. FLUID ELEMENT AS AN OPEN SUBSYSTEM

When the fluid is treated as a continuous medium, the fluid element should be defined to have the following physical and mathematical properties. First, the volume element should be a local thermodynamic equilibrium subsystem containing a large numbers of fluid molecules, so that an instrument lying anywhere but with the same volume would give a measurement of the local thermodynamic quantities such as density, temperature, and thermodynamic pressure; Second, the volume element should be an object containing definite matter, so that the kinematic quantities defined on it, such as velocity, acceleration, are real; Third, the volume element should be the minimum element of the mathematical representation space of fluid, to which a coordinate system can be assigned such that all physical quantities defined on it become functions varying with coordinates.

Here a fluid element is assumed to be an open subsystem of fluid molecules instead of a closed one (which has energy exchange but no mass exchange other than molecular diffusion with its environment), whatever it is small from a macroscopic perspective or large from a microscopic perspective. The fluid consists of fluid elements in such a way that it is a finite covering overlay but not a infinite aggregation, because every fluid element is assumed to have a definite volume of fluid in its space-time of definition. Comparing with the conventional description, if we use the center point at a particle paths to identify the same fluid element, it has to be assumed that the fluid element can change its content in an orderly way. Further, we will attach some inner structural features to a fluid element, such as the averaged micro-rotation over it and the relative material orientation with its adjacent/overlapping ones. These features will exhibit how a group of fluid elements with the same appearance are organized to be an integer. In general, they also make the flowing fluid in the space-time equivalent to an 4-dimensional (4D) differentiable manifold, that is to say, the fluid is a collection of pasting together pieces of a basic Euclidean space represented by a fluid element.

In addition to the classical quantities such as velocity $V_i$, density $\rho$, pressure $p$, and temperature $T$, the orientation variance between fluid elements with respect to space and time gives rise to some new fields indicated by an axial-vector-valued differential 1-form.

$$\mathbf{W}^i = W^i_\mu dx_\mu = \Phi^i dt + A^i_k dx_k . \quad (1)$$

It is natural to refer to $\mathbf{W}^i$ as the vortex fields, where the temporal part $\Phi^i$ is referred to as the eddy field and the spatial part $A^i_k$ the swirl field. Mathematically, the vortex fields correspond to the connection on frame bundle of fluid manifold [5], that is

$$D\mathbf{e}_i = \varepsilon_{ijk} \mathbf{W}^k \mathbf{e}_j \quad (2)$$

where $D$ is the covariant exterior differential operator, and $\{p; \mathbf{e}_1, \mathbf{e}_2, \mathbf{e}_3\}$ is the Cartesian frame at space-time point $p$. When the frames come to pass a rotation

$$\mathbf{e}_i' = R_{ij}(\mathbf{x},t)\mathbf{e}_j , \quad (3)$$

the vortex fields have the following transformation relations

$$\mathbf{W}'^k = R_{kl}\mathbf{W}^l + \mathbf{w}^k , \quad \mathbf{w}^k = \frac{1}{2}\varepsilon_{klm} R_{mp} dR_{lp} , \quad (4)$$

where $d$ is the ordinary exterior differential operator, and $\mathbf{W}'^k$ are defined by $D\mathbf{e}_i' = \varepsilon_{ijk}\mathbf{W}'^k \mathbf{e}_j'$. In general, the vortex fields are non-integrable and so can't ascribe to a rotation between the normalized orthogonal frames. The non-integrable topological structures of fluid in space-time resulted from the vortex fields are exactly indicated by the vortex intensities (mathematically the curvature tensor)

$$\mathbf{F}^i = D\mathbf{W}^i = d\mathbf{W}^i + \varepsilon_{ijk}\mathbf{W}^j \wedge \mathbf{W}^k . \quad (5)$$

where the mark $\wedge$ denotes the exterior product functor. The vortex intensities $\mathbf{F}^i$ have an expression of an axial-vector-valued differential 2-form in space-time, and the tensorial transformation relation $\mathbf{F}'^i = R_{ij}\mathbf{F}^j$. Using the area element $ds_i = \frac{1}{2}\varepsilon_{ijk} dx_j \wedge dx_k$, the vortex intensities have expansion

$$\mathbf{F}^i = \mathbf{B}^i + \mathbf{H}^i = B^i_k ds_k + H^i_k dx_k \wedge dt , \quad (6)$$

with

$$\left.\begin{array}{l} B^i_k = \varepsilon_{klm}\left(\partial_l A^i_m + \frac{1}{2}\varepsilon_{ipq} A^p_l A^q_m\right), \\ H^i_k = \partial_k \Phi^i - \partial_t A^i_k + \varepsilon_{ipq} A^p_k \Phi^q . \end{array}\right\} \quad (7)$$

Usually, the spatial part $B^i_k$ is referred to as the swirl intensity and the spatiotemporal part $H^i_k$ the eddy intensity. Exteriorly differentiating Eq. (6) yields the structure equations (mathematically called the Bianchi identities)

$$D\mathbf{F}^i = d\mathbf{F}^i + \varepsilon_{ijk}\mathbf{W}^j \wedge \mathbf{F}^k = 0 . \quad (8)$$

### III. KINEMATICS OF MOTION FIELDS

The eddy field $\Phi^i$ is a kinematic characteristic of fluid other than the velocity field $V_i$, which can be strictly referred to as the ensemble rotational velocity of molecules around multi-centers in a fluid element when the ensemble translational velocity (namely the velocity field) is removed. Because the rotation of molecules in a fluid element are actually carried out in a conglomerating form with dissipation scale, as observed in real turbulent flows, we can always claim that the eddy field is the representation of the micro-rotation in the fluid element. Therefore, two kinematic characteristics for the fluid element are adopted in our model, where the eddy field represents a cluster of micro-rotations while the velocity field represents the global translation.

Naturally, it is necessary to introduce two subsequent properties, namely the (mass) density $\rho$ and the inertia moment density $\hbar$ of micro-rotation, to calculate the momentum and the body (micro-)moment (namely angular momentum) of fluid element. Because the micro-rotation is not an unitary motion, or in other words there can be many permitted micro-rotations in a fluid element, we assume all micro-rotations have the same characteristic scale, and so the micro-rotational inertia density is assumed to be proportional to the mass density: $\hbar = \rho\Lambda_0$, where the parameter $\Lambda_0$ has the measure of area, indicating the characteristic scale of a single micro-eddy. Another sequent problem is the transformation relation of the inertial force and body moment when the frames come to pass a rotation transformation (3).

It is easy to know, in new frames the components of motion fields transform as $V_i' = R_{ij}(\mathbf{x},t)V_j$ and $\Phi'^i = R_{ij}(\mathbf{x},t)\Phi^j + w^i_4$. First we denote the time derivatives of motion fields following the motion of the fluid element by $a_i' = dV_i'/dt$, $a_i = dV_i/dt$, and $\alpha'^i = d\Phi'^i/dt$, $\alpha^i = d\Phi^i/dt$. The inertial force and body moment at the Cartesian frames have expressions

$$f_{\mathrm{in}\,i} = \rho a_i , \quad m^i_{\mathrm{in}} = \hbar \alpha^i \quad (9)$$

If the frames change, the inertial force and body moment should be objective and have the tensorial transformation relations $f_{\mathrm{in}\,i}' = R_{ij}f_{\mathrm{in}\,j}$, $m'^i_{\mathrm{in}} = R_{ij}m^j_{\mathrm{in}}$. That means we need some additional terms if we want to express the inertial force and body moment in terms of $a_i'$ and $\alpha'^i$, such that

$$f_{\text{in}\,i}' = \rho\left(a_i' + \varepsilon_{ijk}\Omega^j V_k'\right), \\ m'^i_{\text{in}} = \hbar\left[\alpha'^i - dw_4^j/dt + \varepsilon_{ijk}\Omega^j\left(\Phi'^k - w_4^k\right)\right]. \quad (10)$$

where $\Omega^i = w_4^j + V_k w_k^j$ representing the frame rotation velocity following the motion of the fluid element. The additional term in inertial force is usually called the Coriolis force. Because the eddy field has opposite meaning to the frame rotation, the frame rotation $w_4^k$ subtracted from the observed eddy field $\Phi'^k$ gives the substantial eddy field, and the Coriolis effect also attributes to an additional term.

An essential difference between the eddy and the velocity is that the motion state represented by the former is highly dissipative and so that its appearance demands to a proportional supporting body moment $\xi = \mu\Phi^i$. This brings forth that the eddy field is someway different from the transferable quantities such as mass, momentum, heat, etc., and gets a dissipation energy term $(1/2)\mu\Phi^i\Phi^i$ in addition. An extreme case is that only the dissipation body element dominates the evolution process when the inertial moment is negligible because the micro-rotational inertia density is very small.

The motion state indicated by the eddy field is highly dissipative and its energy comes from the viscous dissipation process of the velocity field. Therefor, the increase of energy of micro-rotation does't mean the needing of the input of external energy, and instead indicates the ability to maintain the present state. So for the kinetic energy density of a fluid, the external energy needed to input should be

$$f_{\text{in}\,i} V_i - m^i_{\text{in}}\Phi^i = \rho a_i V_i - \hbar\alpha^i\Phi^i. \quad (11)$$

IV. COUPLING MECHANISMS OF FLOW FIELDS

When the fluid flows, many molecules move away from each other because of the shear process. For the curved flow, the molecules change their orientation in the meantime. The ensemble change of the contact relation between molecules, not only the adjacency but also the relative orientation, result in the viscous interaction. That means that a molecule in fluid can't change its position and orientation freely regardless of the process of microscopic diffusion. On other words, just because of the memory effect to the flow history and the boundary geometry, the molecules naturally have a potential to change their orientation during their transporting, which is now indicated by the swirl field $A^i_k$.

The swirl field essentially influences the viscous interaction by coupling with the velocity field, namely the movement of fluid elements occupying different positions. As well known, for the Newtonian fluid the viscous friction has formula $\boldsymbol{\sigma}_i = \mu\partial_j V_i ds_j \wedge dt$. When the swirl field is involved, the partial derivative in the viscous friction will be made use of the covariant one instead of the ordinary one such that

$$\boldsymbol{\sigma}_i = \mu D_j V_i ds_j \wedge dt = \mu\left(\partial_j V_i + \varepsilon_{ilm}A_j^l V_m\right)ds_j \wedge dt. \quad (12)$$

Usually, the set of covariant partial derivatives of the velocity field $D_j V_i$, simply denoted by $Y_{ji}$, is called the shear strength, and has its dissipation energy term $L_Y$. In the case of the Couette flow between two concentric cylinders, new formula (12) yields that there is no viscous force on the hoop section, or in other words, the molecules diffusing and/or adhering along this direction transfer no momentum.

If the swirl intensity does not vanish everywhere, a further effect of the swirl field can result from the molecules diffusing and/or adhering in a loop. According to

$$DD(V_i\mathbf{e}_i dt) = \chi_{ki}\mathbf{e}_i ds_k \wedge dt, \quad \chi_{ki} = \varepsilon_{ilm}B_k^l V_m, \quad (13)$$

a jump of velocity valued by $\chi_{ki}\mathbf{e}_i$ is produced when the loop is formed by the borderline of the area element $ds_k$. The viscous interactions induced by this jump are diverse, as shown in the governing equations (21) and (23), which would be a linear moment when coupling with the velocity field, or a body force when coupling with the swirl intensity, though they all can be educed from the same dissipation energy term $L_\chi$.

In addition to this, the non-homogeneity of the eddy field, indicated by the eddy intensity $H^i_j$ as a result of the eddy field coupling with the swirl field, will induce a diffusion process of micro-rotation, and so have another dissipation energy term $L_H$. As shown in the next section, the energy also does't mean the need of input of external energy but indicates a kind of energy storage of fluid system.

V. LAGRANGIAN DENSITY OF FLUID SYSTEM

Now we can sum up the above mechanisms to conclude a theory of complicate fluid flow. For simplicity, we consider only incompressible and isothermal flows in this paper, and without loss of generality the mass density is assumed to be unit. In order to obtain a compact deduction of the dynamical equations of field variables, the variational calculus on the action of the fluid system $\mathcal{A}[V_i] = \int_\mathcal{D} L[V_i] dv \wedge dt$ will be used, where $\mathcal{D}$ is the space-time domain occupied by fluid, $dv = dx_1 \wedge dx_2 \wedge dx_3$ is the volume element, the Lagrangian density $L$, namely the dissipation energy density of fluid per unit time, is defined by the viscous dissipation minus the works applied by various forces. For instance, the Lagrangian density of the N-S equations can be expressed by $L[V_i] = L_0 + L_{\text{vis}}$, where the basic term $L_0 = (f_{\text{in}\,i} - f_i + \partial_i p)V_i$ is the negative value of works applied by the inertia force $f_{\text{in}\,i}$, the external force $f_i$ and the pressure gradient $\partial_i p$; the viscous dissipation term has the quadratic form $L_{\text{vis}} = (1/2)\mu(\partial_j V_i)(\partial_j V_i)$. When the vortex fields introduced, the viscous dissipation term is essentially expanded. Making reference to some fundamental principles (expatiated below) we can construct the Lagrangian density as follows

$$L\left[V_i; A_k^i, \Phi^i\right] = L_0 + L_Y + L_\chi - L_\Phi - L_H \quad (14)$$

where $L_\Phi = m_{\text{in}}^i\Phi^i + (1/2)\mu\Phi^i\Phi^i$, and $L_Y, L_\chi, L_H$ are non-negative energy functions. In the above construction, the replacement of the covariant derivatives instead of the ordinary ones is called the the minimal replacement principle; the manner to append the dissipation energy terms according to the independent physical processes $\chi_{ki}$, $\Phi^i$ and $H_{ki}$ is called the minimal coupling principle [6,7]. Besides, the negative indications for $L_\Phi$ and $L_H$ lies on that the dissipation processes induced by the micro-rotation also constitute the feedback processes of the small-scale motion that apply work to the large-scale motion.

## VI. BALANCE RELATIONS IN FLUID DYNAMICS

From the Lagrangian density (14), no matter how the detailed formulation it is, we can always make certain the balance relations in fluid dynamics. If the detailed formulation is further given, these balance relations will bring forth the governing equations of field variables including the velocity field $V_i$, the eddy field $\Phi^i$ and the swirl field $A^i_k$.

It is useful to introduce the some generalized forces to indicate the balance relations of fluid dynamics. They are the driving force 4-forms $\mathbf{M}_i$, the viscous friction 3-forms $\boldsymbol{\sigma}_i$, the structural damping 4-forms $\boldsymbol{\Sigma}_i$, the vortex structural moment 2-forms $\mathbf{Q}^i$ (including the swirl moment 2-forms $\mathbf{G}^i$ and the eddy moment flux 2-forms $\mathbf{E}^i$), the vortex source moment 3-forms $\mathbf{J}^i$ (including the dissipation moment 3-forms $\mathbf{m}^i$ and the twisting moment 3-forms $\varepsilon_{ilm}V_l\boldsymbol{\sigma}_m$), with definitions by

$$\mathbf{M}_i = \frac{\partial L_0}{\partial V_i}dv \wedge dt, \quad \boldsymbol{\sigma}_i = \frac{\partial L_Y}{\partial Y_{ki}}ds_k \wedge dt,$$

$$\boldsymbol{\Sigma}_i = \frac{\partial L_\chi}{\partial V_i}dv \wedge dt = \varepsilon_{ilm}\boldsymbol{\Pi}_l \wedge \mathbf{B}^m, \quad \boldsymbol{\Pi}_l = \frac{\partial L_\chi}{\partial \chi_{ki}}dx_k \wedge dt,$$

$$\mathbf{Q}^i = \frac{\partial L_H}{\partial H^i_k}ds_k - \frac{\partial L_\chi}{\partial B^i_k}dx_k \wedge dt = \mathbf{E}^i - \mathbf{G}^i, \quad \mathbf{G}^i = \varepsilon_{ilm}V_l\boldsymbol{\Pi}_m,$$

$$\mathbf{J}^i = \frac{\partial L_\Phi}{\partial \Phi^i}dv + \frac{\partial L_Y}{\partial A^i_k}ds_k \wedge dt = \mathbf{m}^i + \varepsilon_{ilm}V_l\boldsymbol{\sigma}_m, \quad \mathbf{m}^i = \mathbf{m}^i_{\text{in}} + \boldsymbol{\xi}^i.$$

Then introduce the Lagrangian density (14) and denote the variations of field variables $\{V_i; A^i_k, \Phi^i\}$ by $\{\phi_i; \eta^i_k, \eta^i_4\}$, namely

$$\phi_i = \delta V_i, \quad \eta^i_k = \delta A^i_k, \quad \eta^i_4 = \delta \Phi^i,$$

the definition of the variation of the action yields

$$\delta \mathcal{A}[V_i] = \int_\mathcal{D} \delta L[V_i]dv \wedge dt \qquad (15)$$

with

$$\delta L = (M_i + \Sigma_i)\phi_i + \sigma_{ki}\delta Y_{ki} - m^i\eta^i_4 + G^i_k\delta B^i_k - E^i_k\delta H^i_k.$$

Because of

$$\delta Y_{ki} = \partial_k \phi_i + \varepsilon_{ipq}A^p_k\phi_q + \varepsilon_{ipq}\eta^p_k V_q,$$

$$\delta B^i_k = \varepsilon_{klm}\left(\partial_l \eta^i_m + \varepsilon_{ipq}\eta^p_l A^q_m\right),$$

$$\delta H^i_k = \partial_k \eta^i_4 - \partial_t \eta^i_k + \varepsilon_{ipq}\eta^p_k \Phi^q + \varepsilon_{ipq}A^p_k \eta^q_4,$$

we have

$$\delta L = \left(M_i + \Sigma_i - \partial_k \sigma_{ki} - \varepsilon_{ilm}A^l_k\sigma_{km}\right)\phi_i + \partial_k\left(\sigma_{ki}\phi_i\right)$$
$$- \left(m^i - \partial_k E^i_k - \varepsilon_{ipq}A^p_k E^q_k\right)\eta^i_4 - \partial_k\left(E^i_k\eta^i_4\right)$$
$$+ \left[J^i_k - \partial_t E^i_k - \varepsilon_{ipq}\Phi^p E^q_k + \varepsilon_{klm}\left(\partial_l G^i_m + \varepsilon_{ipq}A^p_l G^q_m\right)\right]$$
$$+ \partial_t\left(E^i_k\eta^i_k\right) - \partial_k\left(\varepsilon_{klm}G^i_l\eta^i_m\right).$$

Noting that the variations of field variables are independent to each other and vanish on the boundary, the minimal action principle of fluid system results in the balance relations, written in the compact form as

$$\mathbf{M}_i + \boldsymbol{\Sigma}_i = d\boldsymbol{\sigma}_i + \varepsilon_{ilm}\mathbf{A}^l \wedge \boldsymbol{\sigma}_m = D\boldsymbol{\sigma}_i, \qquad (16)$$

$$\mathbf{J}^i = d\mathbf{Q}^i + \varepsilon_{ipq}\mathbf{W}^p \wedge \mathbf{Q}^q = D\mathbf{Q}^i. \qquad (17)$$

It is obvious in view of the dimension analysis that the Eqs. (17) are of the balance of force moments while the Eqs. (16) are of the balance of forces. Since the dynamical equations (17) about the force moment balance hold on the lower 3D dimensions, there exists an integrability condition between them. Exteriorly differentiating (17) gives

$$D\mathbf{J}^i = -\varepsilon_{ipq}\mathbf{F}^p \wedge \mathbf{Q}^q. \qquad (18)$$

This equation indicates the balance of various vortex moments of force on an 4D space-time volume element. However, if the vortex intensities $\mathbf{F}^i$ and the corresponding generalized forces $\mathbf{Q}^i$ are coaxial, which would be assumed to be always true as pointed out later, the right of (18) identically vanishes. This together with the coaxiality of $L_Y$, namely $\varepsilon_{ipq}Y_{kp}\sigma_{kq} = 0$, yield an important relation between the micro-rotation and the viscous friction

$$D\mathbf{J}^i = 0 \iff D_t m^i = \varepsilon_{ipq}V_p D_k \sigma_{kq} \qquad (19)$$

This shows the fundamental mechanism in fluid dynamics that the dissipation moment of force corresponding to micro-eddies is made out of the inconsistency of the velocity and the volume resultant of viscous friction. However, the coaxiality of the generalized forces and fluxes can be understood as a fundamental constitutive restriction on the viscous interactions of fluid system, indicating the isotropic response property of fluids.

## VII. GOVERNING EQUATIONS OF FIELD VARIABLES

The usual situation in physics is that the viscous terms in the Lagrangian density have the quadratic form in terms of the generalized fluxes or forces, for example the Lagrangian density of the N-S equations is of this type. So it is meaningful to discuss in detail the following quadratic Lagrangian density recalling the dimensional argument

$$L = L_0 - \Lambda_0 \alpha^i \Phi^i + \frac{\mu}{2}\left(Y_{ki}Y_{ki} - \Phi^i\Phi^i\right) + \frac{\mu\Lambda_1}{2}\left(\chi_{ki}\chi_{ki} - H^i_k H^i_k\right) \quad (20)$$

where μ is the viscosity coefficient and $\Lambda_1$ is another material parameter with measure of area, indicating the characteristic scale of self-organization of micro-rotations over the fluid element, and certainly larger than $\Lambda_0$. By the Lagrangian

density (20), the governing equations of the velocity field and the vortex fields are given as

$$\frac{dV_i}{dt} - f_i + \partial_i p = \mu\left(\partial_k Y_{ki} + \varepsilon_{ilm} A_k^l Y_{km}\right) - \mu\Lambda_1 \varepsilon_{ilm}\chi_{kl} B_k^m, \quad (21)$$

$$\Lambda_0 \frac{d\Phi^i}{dt} + \mu\Phi^i = \mu\Lambda_1\left(\partial_k H_k^i + \varepsilon_{ilm} A_k^l H_k^m\right), \quad (22)$$

$$\varepsilon_{ipq}V_p\sigma_{kq} = \mu\Lambda_1\left(\partial_t H_k^i + \varepsilon_{ilm}\Phi^p H_k^q\right) \\ -\mu\Lambda_1\varepsilon_{klm}\left[\partial_l\left(\varepsilon_{ipq}V_p\chi_{mq}\right) + \varepsilon_{irs}A_l^r\left(\varepsilon_{spq}V_p\chi_{mq}\right)\right]. \quad (23)$$

These equations together with the continuity equation $\partial_i V_i = 0$ consist of a completely closed set of equations for the pressure $p$, the velocity $V_i$, the swirl $A_k^i$ and the eddy $\Phi^i$.

The difference between the governing equations of the velocity field with the N-S equations is obvious. Besides the replacement of $\partial_j V_i$ by $Y_{ji}$, the coupling of the swirl field also makes an additional term in the resultant of the viscous friction. Another remarkable difference is the structural damping term $\mu\Lambda_1\varepsilon_{ilm}\chi_{kl}B_k^m$, which is absolutely a new mechanism for the velocity evolution. It includes the parameter $\Lambda_1$, and begins to work just when the swirl intensity, namely the topological singularity of material orientation, appears.

It is remarkable that the eddy field is also controlled by the convection-diffusion process, but the eddy viscosity $(\Lambda_1/\Lambda_0)\mu$ is much larger than $\mu$. If all vortex structures break up and the eddy becomes homogeneous, the vanish of dissipation moment $m^i$ results in an exponential decay of the eddy field. On the other hand, the governing equations of the swirl field have no term other than the viscous interaction. Therefore, the viscosity coefficient $\mu$ takes no effect to the evolution of the swirl field, that means the $Re$ number is no longer a controlling parameter in the cascade of vortex structures. Further, when we clean up the controlling equations (23) according to the controlling variable, we will obtain a set of wave equations for the swirl field, with the local convection velocity as its wave speed.

Except for the case of homogeneous decaying turbulence, the unsteady term relating to $\Lambda_0$ is negligible, and Eqs. (19) becomes

$$\mu\partial_t \Phi^i = \varepsilon_{ipq}V_p D_k\sigma_{kq}. \quad (24)$$

Since $\sigma_{kq}$, $D_k\sigma_{kq}$ represent the viscous friction and its resultant force over the volume element, the Eqs. (23) tell that the disalignment of velocity and viscous friction is the source of evolution of the vortex intensities, while the disalignment of velocity and bulk viscous friction gives rise to the micro-eddies. These equations also show that along the velocity direction the above twisting processes of the viscous friction vanish, and indicate some special evolutionary properties of streamwise vortex. For example, the streamwise vortex may originate from a kinematic effect of 3D rotation, and is easy to maintain in evolution.

## VIII. CONCLUDING REMARKS

Let summarize the main assumptions that a rational theory has put forward to describe the large-scale motion in fluid flow. First, it assumes the existence of a larger and stable sensitive volume to measure the macroscopic properties if some inner orientational structures are permitted. The fluid element with such volume is treated as an open subsystem of fluid molecules, and the fluid is a finite cover of fluid elements. The so-called vortex fields $\{\Phi^i, A_k^i\}$ indicating the orientation variance between the adjacent/overlapping fluid elements, constitute new dynamic fields independent of the velocity field $V_i$. Second, it assumes the coupling mechanism between the swirl field $A_k^i$ and the velocity field to recast the viscous interactions including the viscous friction, and uses the eddy field $\Phi^i$ to describe the structural features of the small-scale motion in fluid flow. The so introduced dynamic processes which are independent to each other appear in the Lagrangian density of fluid system with an elegant and compact form. The minimal variational principle of the action directly yields the balance relations of fluid dynamics. If the detailed formulation of the Lagrangian density is given, the governing equations of field variables are obtained.

With the closeness of the new theory of fluid flow, one naturally asks: What is the advance we can get from it? Let us offer some comments in this regard. First, classical approach deeming the fluid element as a closed subsystem of molecules admits the velocity as the only flow feature. For turbulence, intricate and complicate fluctuations necessitate the statistical viewpoint, but the decomposition of the velocity into the average part and the fluctuation part is too impertinent to catch the structures embeding in turbulence, let alone the derivative mess as closeness problem, etc.. Simply to say, all statistical approaches up to now seems to make us no lack of knowledge but lack of ability to take it.

In this paper, the author's method is very different. A new and rational description of fluid flow is introduced. Here, a strict definition of turbulence can be given, namely turbulence is a kind of flow with nonzero vortex intensities (topological nontrivial). Several noticeable results may be listed as follows. First, when some new field variables are used to describe the vortex state in fluid flow, including the large-scale structure and the small-scale motion, the corresponding governing equations are naturally of the balance of force moments. Second, the twisting processes between the velocity and the viscous friction such as $\varepsilon_{ilm}V_l\sigma_{km}$, $\varepsilon_{ilm}V_lD_k\sigma_{km}$ paly an important role in the vortex evolution, that is to say, not only the shear force but also the movement are necessary for the formation of a vortex state. In addition, the structural damping $\mu\Lambda_1\varepsilon_{ilm}\chi_{kl}B_k^m$ may be the source of intermittent evolution of velocity in turbulence because of the singular existence of swirl intensity $B_k^i$ in the fluid. Finally, there are two characteristic scale parameters $\Lambda_0$, $\Lambda_1$ indicating the fine structure and coherent structure respectively, which distinguish the hierarchies of vortex evolution. That also means that there is a medial hierarchy in which the vortex evolves irrespective of the Re number.

In conclusion, in this paper we have established a rational theory for fluid flow. The whole theoretical constitution could be analogous with the gauge field theory, where the corresponding gauge group is the rotation group and the corresponding gauge field is the vortex field. According to the theory, the complexity of turbulence is accessible to a scale much coarser than that in the conventional view. This practicability promises it a significant potential for further research [8].

ACKNOWLEDGMENT *(HEADING 5)*

Z.W.N. acknowledges the financial support from NSFC (Grants no. 10372038) and the China Postdoctoral Science Foundation, which are greatly appreciated.